# Complementary-Switchable Dual-Mode SHF Scandium-Aluminum Nitride BAW Resonator

Dicheng Mo, Student Member, IEEE, Shaurya Dabas, Student Member, IEEE, Sushant Rassay, Student Member, IEEE, and Roozbeh Tabrizian, Senior Member, IEEE

*Abstract*— This paper presents a bulk acoustic wave (BAW) resonator with complementary switchable operation in the first and second thickness extensional modes (TE$_1$ and TE$_2$) at 7.04 GHz and 13.4 GHz. Two ferroelectric scandium-aluminum nitride (Sc$_{0.28}$Al$_{0.72}$N) layers are alternatively stacked with three molybdenum electrodes, creating a laminated BAW resonator with independent switchability of polarization in constituent transducers. This enables intrinsic switchability of the resonator in TE$_1$ and TE$_2$ modes, when the ferroelectric Sc$_{0.28}$Al$_{0.72}$N layers are poled in the same or opposite directions, respectively. A generalized analytical proof of complementary switchable operation, extended to laminated BAW resonators consisting of arbitrary number of Sc$_x$Al$_{1-x}$N layers, is presented. For the demonstrated prototype, electromechanical coupling coefficients ($k_t^2$) of 10.1% and 10.7%, and quality factors ($Q$) of 115 and 151, are measured for TE$_1$ and TE$_2$ modes, respectively, when the resonator is configured in corresponding operation states. Besides showing intrinsically configurable operation in super-high-frequency regime with high $k_t^2$ and $Q$, laminated Sc$_{0.28}$Al$_{0.72}$N BAW resonator exhibits repeatable operation under switching cycles.

## I. Introduction

With the exponential increase in wireless data traffic, adaptive spectrum allocation becomes increasingly vital to avoid congestion and interference. Realization of adaptive spectrum allocation in wireless systems requires reconfigurable spectral processors that enable dynamic control over pass- and stop-bands at the radio frequency front-end (RFFE) [1-4].

Currently, the integrated RF duplexers and filters are created using aluminum nitride (AlN) surface and bulk acoustic wave (S/BAW) resonators. High quality-factor ($Q$) AlN BAW and SAW filter technologies with frequencies as high as 6 GHz are extensively adopted in RFFE of modern wireless systems [2-6]. These technologies, however, do not provide any intrinsic frequency tunability or switchability, and their operation is limited to a fixed band. Therefore, extension of communication capacity, to enhance data rates and exploit uncongested spectrum in cm- and mm-wave regimes, requires arraying a large set of fixed-frequency filters using external switches to enable band selection and data aggregation. This strategy is not scalable considering the unfavorable increase in RFFE footprint with the addition of new filters, and excessive loss and power consumption of multiplexers needed for switching.

Alternative acoustic resonator technologies were proposed to achieve intrinsic configurability based on the use of perovskite ferroelectric and paraelectric transducers [7-11]. In these technologies the dependence of transducer polarization and acoustic velocity on DC electric field enables intrinsic switching and frequency tuning of the resonator. However, the major limitation of these technologies is their frequency scaling beyond the ultra-high-frequency regime (UHF: 0.3-3 GHz). This is due to the excessive electrical and mechanical loss of conventional perovskite and ferroelectric films and the processing challenges with thickness miniaturization upon extreme frequency scaling of the resonators.

The recent discovery of ferroelectricity in scandium-doped aluminum nitride films (Sc$_x$Al$_{1-x}$N) is poised to surpass the limitations of perovskites and enable intrinsically configurable resonators with extreme scaling to super-high-frequency regime (SHF: 3-30 GHz) [12]. Configurable thickness-mode BAW and intrinsically switchable Lamb-wave resonators with frequencies up to 10 GHz are recently demonstrated using ferroelectric Sc$_x$Al$_{1-x}$N films [13-17]. In this paper, we present a novel SHF BAW resonator architecture based on laminating two Sc$_x$Al$_{1-x}$N transducers with independent electrical control on their polarization direction. This architecture enables complementary-switchable operation of the resonator in different thickness-extensional mode harmonics (TE$_i$, $i = 1,2$) with consistently large electromechanical coupling coefficient ($k_t^2$) and $Q$.

## II. Concept

Efficient excitation of acoustic resonance modes in piezoelectric transducers require harmonic alignment of mechanical stress profile and applied electric field. In conventional thickness-mode BAW resonators, the electric field is applied uniformly across the piezoelectric film thickness through the top and bottom metal electrodes. This architecture

Manuscript received March 10, 2022. This work was supported by the Defense Advanced Research Projects Agency (DARPA), Tunable Ferroelectric Nitrides (TUFEN) Program under Grant HR00112090049.

D. Mo, S. Dabas, S. Rassay and R. Tabrizian are with the Electrical and Computer Engineering Department, University of Florida, Gainesville, FL 32603 USA, (e-mail: dicheng.mo@ufl.edu).

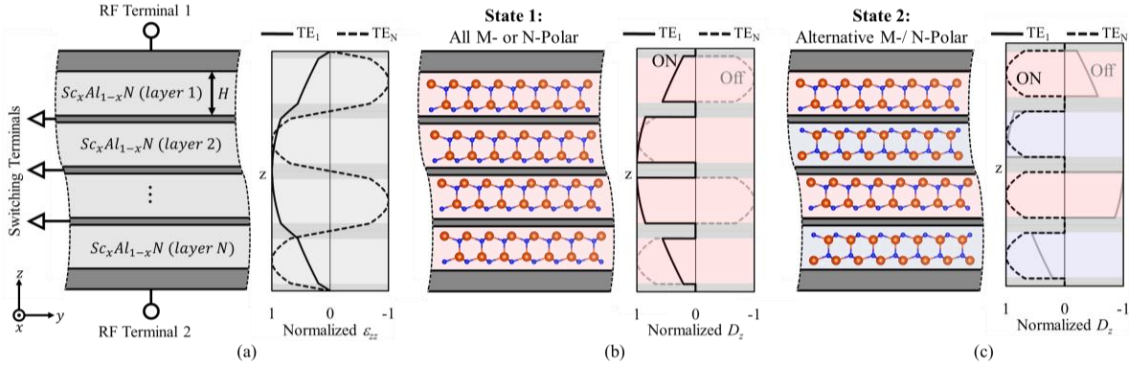

Fig. 1. (a) Schematic of a laminated BAW resonator structure with $N$ $Sc_xAl_{1-x}N$ layers and $N+1$ metal layers, and the normalized z-axis strain for $TE_1$ and $TE_N$ modes for $N = 4$. (b) State 1: the $Sc_xAl_{1-x}N$ layers in the resonator are uniformly M- or N- polar. The $D_z$ is constructive for $TE_1$ and destructive for $TE_N$, resulting in maximum and 0 $k_t^2$ for $TE_1$ and $TE_N$, respectively. (c) State 2: $Sc_xAl_{1-x}N$ layers in the resonator are alternating between M- and N- polar. The $D_z$ is destructive for $TE_1$ and constructive for $TE_N$, resulting in 0 and maximum $k_t^2$ for $TE_1$ and $TE_N$, respectively.

limits the electrically excitable acoustic modes to odd thickness-extensional and thickness-shear harmonics. Further, the $k_t^2$ of the higher-order odd modes decreases, proportional to the mode number, due to the excessive charge cancellation across the transducer thickness.

The laminated BAW resonator is created from alternative stacking of $N$ $Sc_xAl_{1-x}N$ films with $N + 1$ metal layers (integer $N \geq 1$). This structure enables creation of a resonator with a single RF port and $N$ isolated DC ports for independent polarization control of $Sc_xAl_{1-x}N$ films.

Figure 1(a) schematically shows the laminated BAW resonator structure, where the RF electric field is applied uniformly across the $N$ $Sc_xAl_{1-x}N$ films through the top and bottom electrodes, while the intermediate electrodes enable independent control of the polarization direction in $Sc_xAl_{1-x}N$ layers based on the ferroelectric behavior. Figure 1(a) also shows the z-axis strain for the first and $N^{th}$ ($N = 4$) thickness-extensional modes (i.e., $TE_1$ and $TE_N$).

Reversing polarization direction, by applying low-frequency switching pulses [18, 19], enables 180° phase-shift in mechanical excitations. This facilitates excitation of both odd and even thickness modes with similarly large $k_t^2$, despite the uniform electric field across all the $Sc_xAl_{1-x}N$ layers.

Assuming infinitesimally thin metal electrodes, the mode-shape function of the $M^{th}$ thickness-extensional BAW mode ($TE_M$) in the lamination of $N$ $Sc_xAl_{1-x}N$ layers with thickness of $H$ is formulated by the normalized z-axis strain $\varepsilon_{zz,M}(z)$ as:

$$\varepsilon_{zz,M}(z) = \sin\left(\frac{M\pi}{NH}z\right) \quad (1).$$

Considering the linear dependence of the effective longitudinal piezoelectric coefficient ($e_{33,eff}$) in $Sc_xAl_{1-x}N$ on its instantaneous polarization ($P_{inst}$), the electric displacement ($D_{z,M}$) across the laminated stack is derived as:

$$D_{z,M}(z) = e_{33}P_{inst}(z)\varepsilon_{zz,M}(z) \quad (2).$$

Here, $e_{33}$ is the longitudinal piezoelectric constant for a perfectly metal-polar (M-polar) $Sc_xAl_{1-x}N$ film. The motional charge per unit area ($Q_{m,M}$) is derived from:

$$Q_{m,M} = \int_0^{NH} \frac{D_{z,M}(z)}{NH} dz$$
$$= e_{33}\sum_{i=1}^{N}\int_{(i-1)H}^{iH} \frac{P_{inst,i}\varepsilon_{zz,M}(z)}{NH} dz \quad (3).$$

Here, $P_{inst,i}$ is $P_{inst}$ in the $i^{th}$ $Sc_xAl_{1-x}N$ layer. Using Eq. (3), the $k_t^2$ of the $TE_M$ mode can be derived as [20]:

$$k_{t,M}^2 = \frac{\frac{1}{2}\frac{Q_{m,M}^2}{C_0}}{\frac{1}{2}\frac{Q_{m,M}^2}{C_0} + \int_0^{NH} \frac{c_{33}}{2}\varepsilon_{zz,m}^2(z)dz} \quad (4).$$

Here, $c_{33}$ is the z-axis elastic constant of $Sc_xAl_{1-x}N$, and $C_0$ is the laminate capacitance per unit area:

$$C_0 = \frac{\epsilon_{33}}{NH} \quad (5),$$

where $\epsilon_{33}$ is the z-axis dielectric constant. Considering Eq. (3) and Eq. (4), $Q_{m,M}$ and $k_{t,M}^2$ can be maximized by proper switching of constituent $Sc_xAl_{1-x}N$ layers in the laminate to maximally harmonize $\varepsilon_{zz,M}$. Specifically, opting for uniform (i.e., State 1: $P_{inst,i} = \pm 1$ for all $i$s) or alternating (i.e., State 2: $P_{inst,i} = (-1)^i$) poling of $Sc_xAl_{1-x}N$ layers, two optimum operation states is recognized.

### A. Operation State 1: Uniform Polarization

Opting for uniform polarization direction in all $Sc_xAl_{1-x}N$ layers results in the highest $k_t^2$ for the $TE_1$ mode when inserting $P_{inst,i} = \pm 1$ in Eq. (3) and Eq. (4) and derived as:

$$k_{t,1}^2 = \frac{\frac{e_{33}^2}{\epsilon_{33}}\left(\int_0^{NH} \sin\left(\frac{\pi z}{NH}\right) dz\right)^2}{\frac{e_{33}^2}{\epsilon_{33}}\left(\int_0^{NH} \sin\left(\frac{\pi z}{NH}\right) dz\right)^2 + \int_0^{NH} \frac{c_{33}}{2}\left(\sin\left(\frac{\pi z}{NH}\right)\right)^2 dz}$$
$$= \frac{8}{\pi^2}\cdot\frac{e_{33}^2}{\epsilon_{33}c_{33}}\cdot\frac{1}{1+\frac{8}{\pi^2}\cdot\frac{e_{33}^2}{\epsilon_{33}c_{33}}} \quad (6).$$

In this operation state, the motional charge of the $TE_N$ mode ($Q_{m,N}$) is nulled considering:

$$Q_{m,N} = e_{33}\int_0^{NH} \sin\left(\frac{N\pi z}{NH}\right) dz = 0 \quad (7).$$

This results in a $k_t^2$ of 0 for the $TE_N$ mode. Besides the $TE_1$ mode, it should be noted that higher odd harmonics (i.e., $M = 2a + 1$; $a$ is a positive integer) can also be excited in this operation state. However, the $k_t^2$ of these harmonics is significantly lower than the $TE_1$ mode and is derived using Eq. (6) and Eq. (7) as:

$$k_{t,2a+1}^2 = \frac{k_{t,1}^2}{(2a+1)^2} \tag{8}.$$

These harmonics may appear as spurious bands, when the resonator is used for spectral processing in RFFE. Considering the low $k_t^2$ and distant location in frequency spectrum, these spurious bands can be concealed by proper filter design.

### B. Operation State 2: Alternating Polarization

On the other hand, opting for alternating polarization switching of the layers in laminate (i.e., State 2: $P_{inst,i} = (-1)^i$) results in perfectly constructive accumulation of motional charge for the $TE_N$ mode considering:

$$Q_{m,N} = e_{33} \sum_{i=1}^{N} \int_{(i-1)H}^{iH} (-1)^i \sin\left(\frac{\pi}{H}z\right) dz$$

$$= N \int_0^H \sin\left(\frac{\pi}{H}z\right) dz \tag{9}$$

This results in a maximum $k_t^2$ for $TE_N$ mode derived as:

$$k_{t,N}^2 = \frac{\frac{e_{33}^2}{\epsilon_{33}}\left(N \int_0^H \sin\left(\frac{\pi}{H}z\right) dz\right)^2}{\frac{e_{33}^2}{\epsilon_{33}}\left(N \int_0^H \sin\left(\frac{\pi}{H}z\right) dz\right)^2 NH + c_{33} \int_0^{NH} \varepsilon_{zz,N}^2(z) dz}$$

$$= \frac{8}{\pi^2} \cdot \frac{e_{33}^2}{\epsilon_{33}c_{33}} \cdot \frac{1}{1 + \frac{8}{\pi^2} \cdot \frac{e_{33}^2}{\epsilon_{33}c_{33}}} \tag{10}.$$

In this operation state, the motional charge, $Q_{m,1}$, for the $TE_1$ mode is derived as:

$$Q_{m,1} = e_{33} \sum_{i=1}^{N} \int_{(i-1)H}^{iH} (-1)^i \varepsilon_{zz,M}(z) dz =$$

$$\frac{e_{33}}{2} \sum_{i=1}^{N} (-1)^i \left(\int_{(i-1)H}^{iH} \varepsilon_{zz,1}(z) dz + \int_{(N-i)H}^{(N-i+1)H} \varepsilon_{zz,1}(z) dz\right) \tag{11}$$

Replacing $TE_1$ mode z-axis strain $\varepsilon_{zz,1}$ and using auxiliary variable $\hat{z} = NH - z$, Eq. (10) can be simplified using:

$$\int_{(N-i)H}^{(N-i+1)H} \sin\left(\frac{\pi}{NH}z\right) dz \,\&= -\int_{(i-1)H}^{iH} \sin\left(\frac{\pi}{NH}(NH - \hat{z})\right) d\hat{z}$$

$$= -\int_{(i-1)H}^{iH} \sin\left(\frac{\pi}{NH}\hat{z}\right) d\hat{z} \tag{12}.$$

Inserting Eq. (12) in Eq. (11), $Q_{m,1}$ is nulled. This results in a $k_t^2$ of 0 for the $TE_1$ mode, when operating in State 2.

This proves the complementary operation of the resonator in $TE_1$ and $TE_N$ modes, when the stack polarization is configured in uniform (State 1) or alternating (State 2) states, respectively. In this paper, laminated $Sc_xAl_{1-x}N$ BAW resonators with $N = 2$ layers and complementary switchable operation in $TE_1$ and $TE_2$ modes is demonstrated.

It should also be noted that in State 2, besides the $TE_N$ mode, only higher-order harmonics with $M = N(2a + 1)$ can be excited. The $k_t^2$ of these modes is derived using Eq. (6) and Eq. (7) as:

$$k_{t,N(2a+1)}^2 = \frac{k_{t,N}^2}{(2a+1)^2} \tag{13}.$$

The complementary operation principle is schematically demonstrated in Fig. 1(b-c), showing the electric displacement for $TE_1$ and $TE_N$ ($N = 4$) modes in the two operation states. In State 1 (Fig. 1(b)), all the $Sc_xAl_{1-x}N$ layers are all either metal- (i.e., M-polar: Sc or Al) or nitrogen-polar (i.e., N-polar), resulting in transduction of the $TE_1$ and suppression of $TE_N$ modes. In State 2 (Fig. 1(c)), the $Sc_xAl_{1-x}N$ layers are alternating between M- and N-polar resulting in suppression of $TE_1$ and transduction of $TE_N$ modes.

### III. RESONATOR MODELING AND DESIGN

In presence of electrodes with finite thicknesses, the operation of laminated $Sc_xAl_{1-x}N$ BAW resonator can be modeled using the Mason's waveguide approach [21]. Figure 2(a) schematically shows the cross-section of the resonator reported in this paper, consisting of two 150nm-thick $Sc_xAl_{1-x}N$ layers, three 50nm-thick molybdenum (Mo) electrodes, stacked atop of 58.5nm-thick AlN seed layer. Figure 2(b) shows the resonator Mason's model consisting of cascaded waveguides representing the laminate layers. In this model $Z_{s,i}$ and $Z_{p,i}$ are the series and shunt acoustic impedance of the layer $i$ ($i \in \{Mo, ScAlN, AlN\}$) in the laminate and formulated as:

$$Z_{s,i} = jZ_i \tan(\beta_i d_i/2),$$
$$Z_{p,i} = -jZ_i/\sin(\beta_i d_i) \quad (14).$$

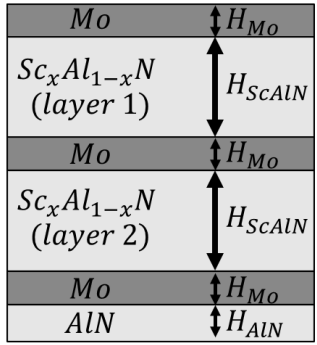
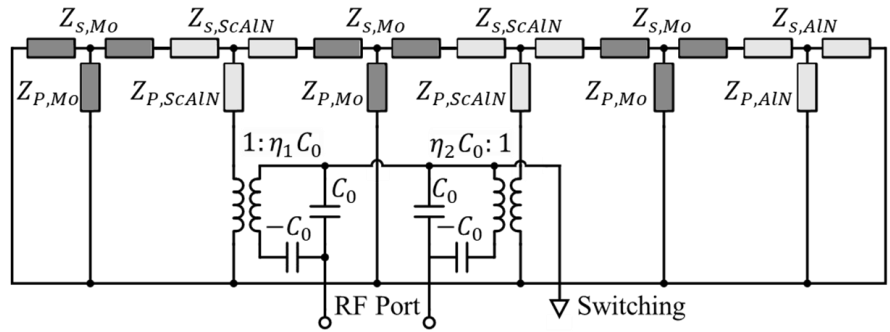

(a)                                                             (b)

Fig. 2. (a) Cross-section schematic of the proposed laminated BAW resonator highlighting constituent layers and their thickness. (b) The Mason's waveguide model for the laminated BAW resonator corresponding to the stack shown in Fig. 2(a).

TABLE 1.
MATERIAL PROPERTIES AND THICKNESSES OF LAYERS
IN LAMINATED $Sc_xAl_{1-x}N$ BAW RESONATOR, USED FOR MODELING.

| Parameter | Mo | Al | $Sc_{0.28}Al_{0.72}N$ | AlN |
|---|---|---|---|---|
| Layer Thickness (nm) | 50 | 50 | 145 | 58 |
| Acoustic Impedance ($10^6$ Pa·s/m) | 66.38 | 17.39 | 31.27 | 35.83 |
| BAW Velocity (m/s) | 6,508 | 6,441 | 8,808 | 10,857 |
| $e_{33}$ (C/m$^2$) | - | - | 2.15 | 1.55 |
| Relative Permittivity | - | - | 20.6 | 9 |

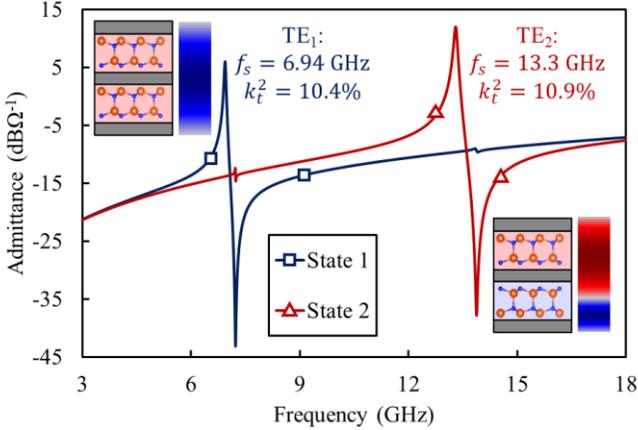

Fig. 3. Simulated admittance of the laminated resonator, using Mason's model. In State 1 the $Sc_{0.28}Al_{0.72}N$ layers are uniformly poled, resulting in transduction of TE$_1$ and suppression of TE$_2$. In State 2 the $Sc_{0.28}Al_{0.72}N$ layers are poled in opposite direction, resulting in transduction of TE$_2$ and suppression of TE$_1$.

Here, $Z_i$, $\beta_i$ and $d_i$ are the acoustic impedance per unit area, the wavenumber, and the thickness of corresponding layers, defined by $z$-axis elastic constant ($c_{33,i}$) and mass-density ($\rho_i$) of each layer and the operation frequency ($f$) as:

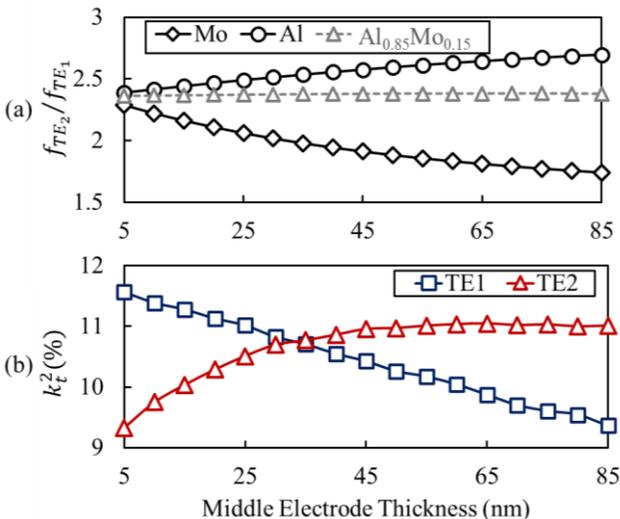

Fig. 4. (a) Simulated frequency ratio of TE$_1$ and TE$_2$ modes for different center electrode materials and thicknesses. (b) Simulated $k_t^2$ of TE$_1$ and TE$_2$ modes for different thicknesses of Mo center-electrode.

$$Z_i = c_{33,i}\beta_i/(2\pi f),$$
$$\beta_i = (2\pi f)/\sqrt{c_{33,i}/\rho_i} \quad (15).$$

The configurable operation of the resonator is modeled using the polarization-dependent piezoelectric coefficient (i.e., $e_{33,eff,i} = e_{33}P_{inst,i}$) for the transformers representing electromechanical transduction ($\eta_i$) and in $Sc_xAl_{1-x}N$ layers as:

$$\eta_i = e_{33}P_{inst,i}/\epsilon_{33} \quad (16).$$

Finally, the static capacitors per unit area corresponding to each $Sc_xAl_{1-x}N$ layers is derived as:

$$C_0 = \epsilon_{33}/d_i \quad (17).$$

Figure 3 shows the simulated admittance of the resonator, using the material properties and layer thicknesses listed in Table 1 for the constituent layers, assuming a top-electrode area of 6,500μm$^2$. In State 1 (i.e., $P_{inst,1} = P_{inst,2}$), TE$_1$ is excited at 6.94 GHz, with a $k_t^2$ of 10.4%, while TE$_2$ is suppressed. Alternatively, in State 2 (i.e., $P_{inst,1} = -P_{inst,2}$), TE$_2$ is excited at 13.3 GHz with a $k_t^2$ of 10.9%, while TE$_1$ is suppressed.

The frequency ratio of TE$_1$ and TE$_2$ modes for this structure is 1.92. This ratio can be accurately controlled by changing the thickness or material of the middle electrode layer. Figure 4(a) shows the TE$_1$ and TE$_2$ frequency ratio as a function of the middle electrode thickness for aluminum (Al) and Mo. When using Mo, the thickness of the middle electrode can be optimized to achieve an integer ratio of 2. Having a resonator with complementary-switchable modes with integer ratio is highly desirable for sub- and super-harmonic communication schemes that are used for enhancement of data rate and reduction of latency in congested wireless networks [22, 23]. Further, using Mo-Al alloy, with 15% Mo atomic fraction, for the middle electrode enables sustaining a constant frequency ratio across large thickness variations that may result from process uncertainties. Finally, it should be noted that the change in middle electrode thickness has an opposite effect on the $k_t^2$ of TE$_1$ and TE$_2$ modes, considering the contrasted z-axis strain profile of the two modes at the center of the laminate. Figure 4(b) shows the $k_t^2$ of TE$_1$ and TE$_2$ modes for different thicknesses of Mo middle electrode. Using proper thickness (34nm), the same $k_t^2$ of 10.8% can be achieved for both modes.

## IV. DEVICE FABRICATION

Figure 5 shows the fabrication process for creation of laminated $Sc_{0.28}Al_{0.72}N$ BAW resonators. The process consists of successive deposition and patterning of $Sc_{0.28}Al_{0.72}N$ and Mo layers on a silicon (Si) substrate. $Sc_{0.28}Al_{0.72}N$ are deposited using reactive magnetron sputtering from segmented scandium-aluminum targets [18]. Mo layers are deposited using DC sputtering. Prior to deposition of the first Mo layer, a ~58nm AlN layer is deposited to serve as the seed, enabling (110)-textured growth of Mo film [24]. This further helps c-axis textured-growth of $Sc_{0.28}Al_{0.72}N$ films.

Figure 6(a) shows the cross-sectional scanning electron microscope (SEM) image of the stack, highlighting the thickness of constituent layers in the laminate. Prior to deposition of the first $Sc_{0.28}Al_{0.72}N$ layer, the bottom Mo is patterned using boron trichloride (BCl$_3$) gas in an inductively coupled plasma reactive-ion-etching (RIE ICP) system. The bottom Mo electrode is patterned using tapered photoresist mask features created by proximity exposure mode photolithography. This enables formation of Mo patterns with slanted sidewalls, which is essential for crack-free growth of successive $Sc_{0.28}Al_{0.72}N$ layers. A similar procedure is repeated to pattern middle Mo electrodes and prior to deposition of the second $Sc_{0.28}Al_{0.72}N$ layer. Figure 6(b) shows the cross-sectional SEM of the transducer stack at the edge of patterned bottom and middle Mo electrodes, highlighting the slanted sidewall and crack-free texture of $Sc_{0.28}Al_{0.72}N$ layers.

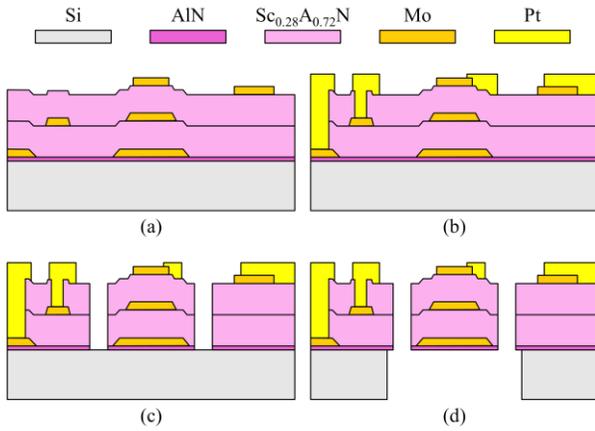

Fig. 5. Fabrication process for creation of laminated $Sc_{0.28}Al_{0.72}N$ BAW resonators: (a) Deposition and patterning of transducer. (b) Etching access to bottom and metal electrodes and metallization. (c) Etching trenches to define lateral geometry. (d) Releasing device from backside.

To evaluate the crystal quality and degree of *c*-axis orientation, X-ray omega scans (*i.e.*, rocking curves) are performed after deposition of each $Sc_{0.28}Al_{0.72}N$ layer on patterned Mo electrodes. Full-width-half-maximums (FWHM) of 2.33° and 2.64° are measured for (002) $Sc_{0.28}Al_{0.72}N$ after deposition of the first and second $Sc_{0.28}Al_{0.72}N$ layers, respectively. The higher FWHM measured after the deposition of second layer may correspond to slight shift in diffraction peak angle due to the different residual stress in top and bottom $Sc_{0.28}Al_{0.72}N$ films.

After deposition of the second $Sc_{0.28}Al_{0.72}N$ layer and the third Mo layer, the top electrodes are patterned using a similar etch process as bottom and middle Mo layers, but with photoresist mask created in contact mode lithography, to achieve straight walls.

Following deposition of the transducer stack, the $Sc_{0.28}Al_{0.72}N$ layers are etched using a timed chlorine-hydrogen ($Cl_2$-$H_2$) based recipe in RIE ICP system to reach bottom and middle Mo electrodes. Next, 500nm-thick platinum (Pt) layer is deposited and patterned using lift-off to create routing lines and pads. The resonator lateral geometry is then patterned using silicon-dioxide hard-mask and $BCl_3$ recipe in ICP RIE. Finally, the resonators are released by deep-reactive-ion-etching of Si handle layer from the backside of the substrate. Figure 6(c) shows the SEM image of the laminated $Sc_{0.28}Al_{0.72}N$ BAW resonator. The RF and switching terminals are highlighted.

## V. CHARACTERIZATION

Laminated $Sc_{0.28}Al_{0.72}N$ BAW resonators are measured to characterize their switching behavior and RF characteristics. Ferroelectric polarization hysteresis and switching behavior of the constituent $Sc_{0.28}Al_{0.72}N$ layers are characterized using a Radiant PiezoMEMS ferroelectric tester. Resonator RF performance is measured using Keysight N5222A PNA vector network analyzer (VNA), along with short-open-load-through (SOLT) calibration procedure enabled by the designated kits. The resonator nonlinearity is measured using two Keysight N5173B signal generators for two-tone input signal excitation and a Keysight N9010A signal analyzer, with a 20dB ± 1.2dB attenuator, for harmonic detection. Finally, the temperature characteristic of frequency is extracted by measuring resonator in a temperature-controlled Semi-Probe PSL4 RF probe station.

### A. Ferroelectric Characterization

The polarization hysteresis loop of the 150nm-thick $Sc_{0.28}Al_{0.72}N$ layers in the laminate is measured by driving the ferroelectric film using 20kHz bipolar triangular signals with 80V amplitude and measuring instantaneous current. Figure 7(a) shows the measured polarization-voltage hysteresis loop for the top $Sc_{0.28}Al_{0.72}N$ layer, used for resonator switching between $TE_1$ and $TE_2$ modes. A coercive voltage of 79V is measured, highlighting the large voltage required for polarization switching of the film. The polarization and coercive voltage are on par with previous report for $Sc_xAl_{1-x}N$ film with similar Sc content and compressive residual stress [12]. Figure 7(b) shows the measured instantaneous current for the bipolar triangular pulse-train drive that is used for polarization switching between M- and N-polar states. The large current at the transition between positive and negative triangular pulses is induced by the polarization switching. While the application of 80V enables switching with a single pulse, opting for smaller voltages results in partial switching and facilitates observation of resonator response evolution as transitioning between the two states.

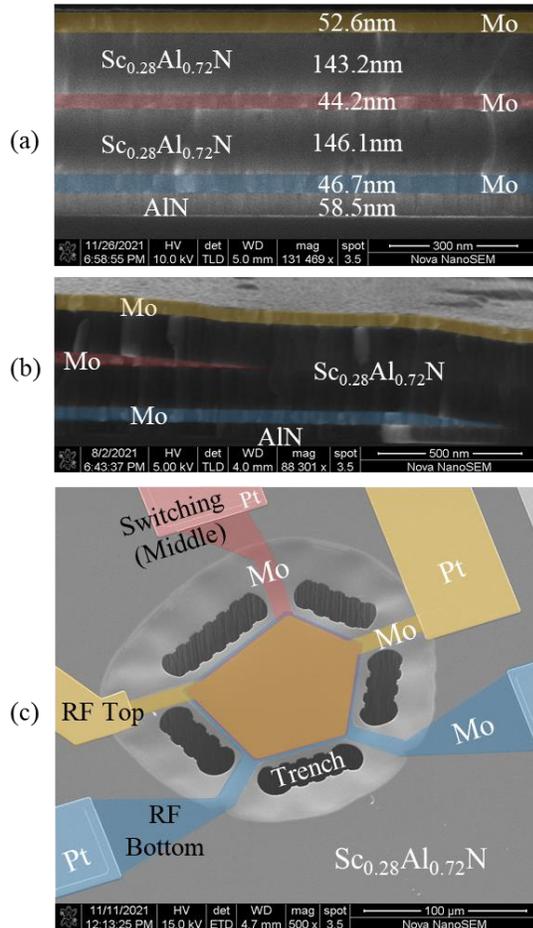

Fig. 6. (a) Cross-sectional SEM of (a) the laminated $Sc_{0.28}Al_{0.72}N$ transducer and (b) at the edge of patterned electrodes, highlighting the smooth slanted sidewalls and crack-free $Sc_{0.28}Al_{0.72}N$ growth. (c) SEM image of complementary switchable $Sc_{0.28}Al_{0.72}N$ BAW resonator.

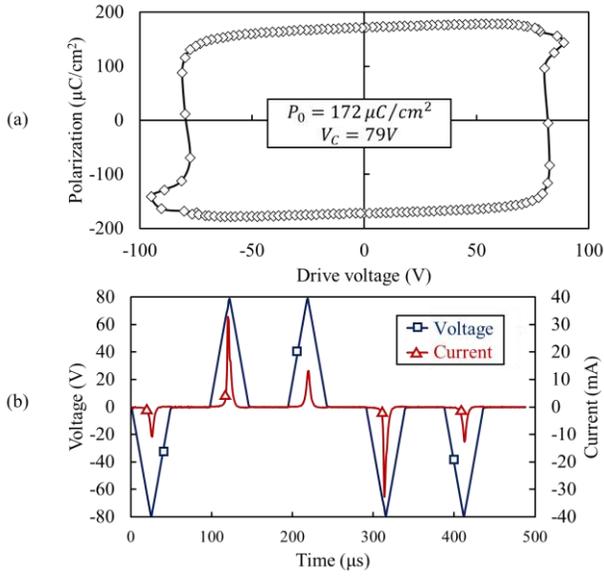

Fig. 7. (a) Measured polarization-voltage hysteresis loop for the top $Sc_{0.28}Al_{0.72}N$ layer, used for resonator switching, highlighting a coercive voltage of 79V. (b) The polarization switching procedure using triangular voltage pulses, and the corresponding polarization switching current, peaking at pulse-sign transitions.

### B. Admittance Measurement

The resonator admittance is extracted from the measured reflection coefficient (*i.e.*, $S_{11}$). Figure 8(a) shows the large-span admittance of the laminated $Sc_{0.28}Al_{0.72}N$ BAW resonator in two operation states. In State 1, both $Sc_{0.28}Al_{0.72}N$ layers are poled in the same direction (M-polar in Fig. 8(a)), resulting in excitation of $TE_1$ and suppression of $TE_2$ modes. Upon application of 20kHz triangular switching pulses to the bottom $Sc_{0.28}Al_{0.72}N$ layer, the resonator reconfigures to State 2, where the $TE_1$ mode is suppressed and $TE_2$ emerges. Figure 8(b-c) show the short-span admittance when operating in each state, highlighting the evolution across inter-state transition. Using ~78V switching voltage, the transition is obtained by application of 5 pulses. Figure 8(b-c) insets show the performance metrics of $TE_1$ and $TE_2$ modes, in State 1 and State 2, respectively. $k_t^2$ and $Q$ values are using [25, 26]:

$$k_t^2 = \frac{\pi^2}{8}\left(\frac{f_p^2 - f_s^2}{f_s^2}\right), Q = \frac{f}{2}\left|\frac{\partial \varphi_Y}{\partial f}\right| \quad (18).$$

Here $f_s$ and $f_p$ are the series and parallel resonance frequencies and $\varphi_Y$ is admittance phase. In State 1, $TE_1$ operates at 7.04 GHz with a $Q$ of 115 and $k_t^2$ of 10.1%. In State 2, $TE_2$ operates at 13.4 GHz with a $Q$ of 151 and $k_t^2$ of 10.7%.

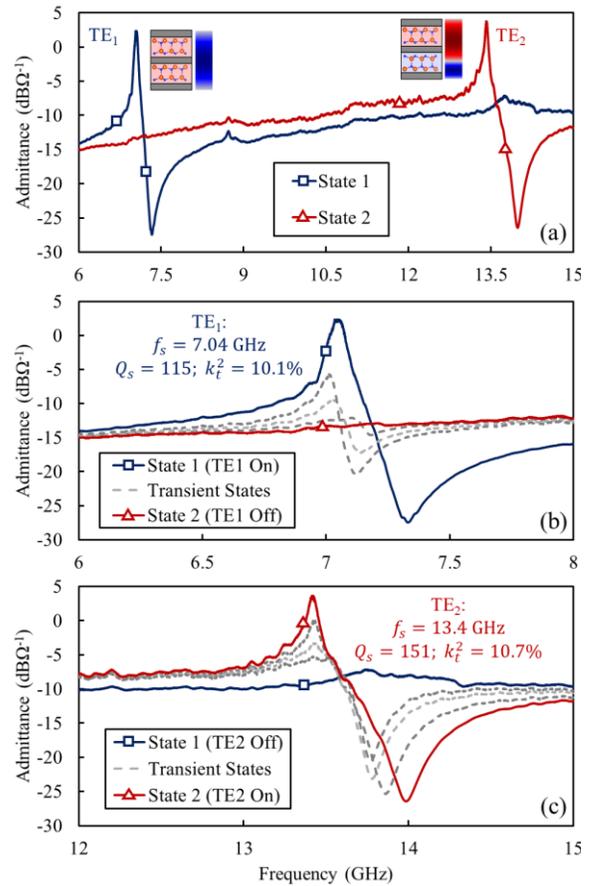

Fig. 8. (a) Measured large-span admittance of the complementary switchable SHF $Sc_{0.28}Al_{0.72}N$ BAW resonator in two operation states. State 1 (uniform polarization of $Sc_{0.28}Al_{0.72}N$ layers) the $TE_1$ is on and $TE_2$ is off. Switching to State 2 (alternating polarization of $Sc_{0.28}Al_{0.72}N$ layers) turns off the $TE_1$ mode and turns on the $TE_2$ mode. Measured short-span admittance of (a) $TE_1$ and (b) $TE_2$ modes, as evolving from State 1 to State 2 by application of switching pulse.

### C. Retention under Switching

The performance retention of the resonator is measured under 50 switching cycles. Figure 9 shows the variations in frequency, $Q$ and $k_t^2$ of the $TE_1$ mode over switching. Early in the switching cycles (*i.e.*, after the first 5 cycles), a slight drop is observed in frequency and $k_t^2$. This may correspond to fractional domain wall pinning and due to defects and imperfect intergranular boundaries [27, 28], which induce undesirable charge cancellation in $Sc_{0.28}Al_{0.72}N$ transducer. After these initial degradations, a stable operation is observed with frequency fluctuations within 0.11%, and $Q$ and $k_t^2$ variations within 7.4% and 3.7%, respectively. Similar to other ferroelectric devices, the performance retention in switchable $Sc_{0.28}Al_{0.72}N$ BAW resonators can be significantly improved by enhancing crystallinity of the film and reduction of defects.

### D. Nonlinearity and Temperature Characteristics

The nonlinearity of laminated $Sc_{0.28}Al_{0.72}N$ BAW resonator is measured using two tones with 10MHz frequency offset. Figure 10(a) shows the nonlinearity characteristic of the resonator, highlighting second- and third-order input intercept point (*i.e.*, IIP2 and IIP3) for the $TE_1$ mode of 59.5 dBm and 40.5 dBm, respectively. These values are comparable with UHF AlN BAW counterparts [29, 30]. The apparent lower slope for intermodulation (IM) at lower end of the input power corresponds to the small magnitude of the third harmonic that is nearly buried by the noise floor of measurement setup. Figure 10(b) shows the measured temperature characteristics of $TE_1$ and $TE_2$ modes frequency, over –20°C to 100°C. Linear temperature coefficient of frequency (TCF) of –43.49 ppm/°C and –41.86 ppm/°C are measured for $TE_1$ and $TE_2$ modes, respectively. The slightly higher temperature sensitivity of $TE_1$ mode may correspond to the placement of middle Mo in high-strain region of the mode-shape and the large negative temperature coefficient of elasticity of Mo [31].

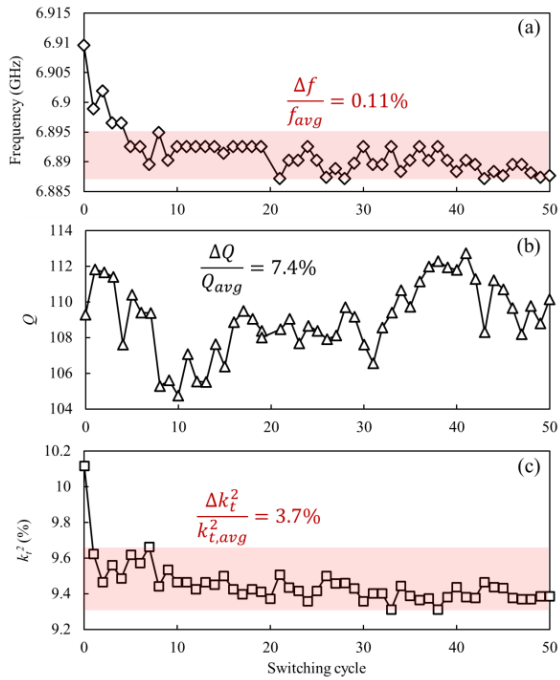

Fig. 9. Measured retention characteristic of $Sc_{0.28}Al_{0.72}N$ BAW resonator's (a) frequency, (b) $Q$, and (c) $k_t^2$, over 50 switching cycles.

## VI. CONCLUSION

This paper presented a new $Sc_xAl_{1-x}N$ BAW resonator technology with complementary-switchable operation in fundamental and high-order thickness extensional modes (i.e., $TE_1$ and $TE_N$). The resonator structure consists of $N$ ferroelectric $Sc_xAl_{1-x}N$ films alternatively stacked with $N+1$ metal electrodes, enabling independent polarization switching of each piezoelectric film. Opting for uniform or alternating poling of the $Sc_xAl_{1-x}N$ layers, the resonator can be switched to operate in two complementary states with either $TE_1$ or $TE_N$ active resonance modes of similarly large $k_t^2$. The generalized principle of complementary-switchable operation in laminated $Sc_xAl_{1-x}N$ BAW resonators were formulated and the role of ferroelectric polarization tuning in excitation and suppression of fundamental and harmonic BAW modes were discussed. Resonator prototype implemented in a laminate consisting of two 150nm-thick $Sc_{0.28}Al_{0.72}N$ layers and three ~50nm Mo layers was presented. The resonator was switched by application of 20kHz triangular pulses to either operate in the $TE_1$ mode, at 7.04GHz with a $Q$ of 115 and $k_t^2$ of 10.1%, or in the $TE_2$ mode, at 13.4 GHz with a $Q$ of 151 and $k_t^2$ of 10.7%. The performance retention of the resonator under 50 switching cycles were presented highlighting repeatable operation after a slight degradation in frequency and $k_t^2$ in the initial cycles. The resonator nonlinearity, when operating in $TE_1$ mode, was measured showing IIP2 and IIP3 of 59.55dBm and 40.50dBm, respectively. The temperature characteristic of $TE_1$ and $TE_2$ modes was measured showing a TCF of –43.49 ppm/°C and –41.86 ppm/°C, respectively. The presented complementary-switchable super-high-frequency BAW resonator was well-suited for creation of intrinsically configurable spectral processors in modern wireless systems and congested networks.

## ACKNOWLEDGMENT

The authors would like to thank the University of Florida Nanoscale Research Facility cleanroom staff for fabrication support and Troy Tharpe for the help with crystal structure modeling.

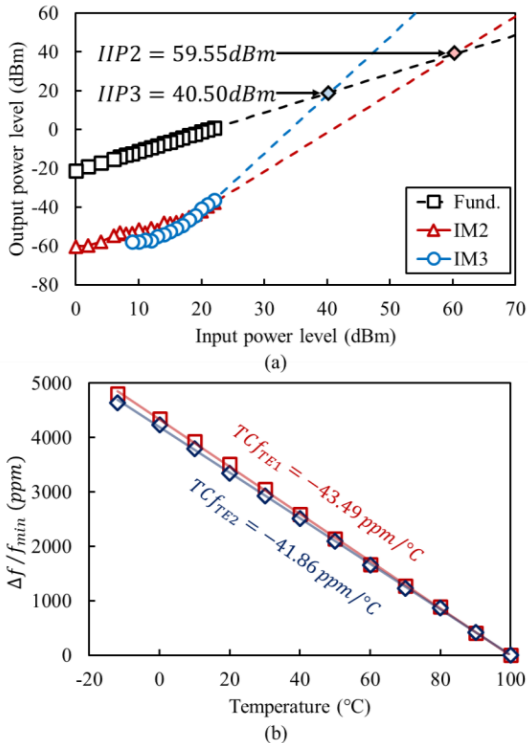

Fig. 10. (a) Measured nonlinearity characteristic of the $TE_1$ mode. The output power is measured with a 20dB ± 1.2dB attenuator to protect the spectrum analyzer. (b) Measured temperature characteristic of $TE_1$ and $TE_2$ modes frequency.